\documentclass[twocolumn,switch, twoside]{article}

\usepackage{arxiv}
\usepackage[utf8]{inputenc} 
\usepackage[T1]{fontenc}    
\usepackage{hyperref}       
\usepackage{url}            
\usepackage{booktabs}       
\usepackage{amsfonts}       
\usepackage{amsmath}
\usepackage{nicefrac}       
\usepackage{microtype}      
\usepackage{lipsum}
\usepackage{fancyhdr}       
\usepackage{graphicx}       
\usepackage{algorithmic}
\usepackage{algorithm}
\usepackage{subcaption}
\graphicspath{{media/}}     
\usepackage[numbers]{natbib}
\usepackage{breakurl}  
\hypersetup{breaklinks=true}  

\AtBeginDocument{
  \newgeometry{
    textheight=9in,
    textwidth=7.5in,
    top=1in,
    headheight=14pt,
    headsep=25pt,
    footskip=30pt,
    centering  
  }
}

\title{Centrally Coordinated Multi-Agent Reinforcement Learning for Power Grid Topology Control}
\begin{document}

\twocolumn[\begin{@twocolumnfalse}

\author{Barbera de Mol\\
EDMij B.V.\\
Ouderkerk a.d. Amstel, The Netherlands\\
\texttt{barbera.de.mol@edmij.nl}
   \And
Davide Barbieri\\
TenneT TSO B.V.  \\
Arnhem, The Netherlands\\
\texttt{davide.barbieri@tennet.eu}
   \AND
Jan Viebahn\\
TenneT TSO B.V.\\
Arnhem, The Netherlands \\
\texttt{jan.viebahn@tennet.eu}
   \And
Davide Grossi\\
University of Groningen\\
Groningen, The Netherlands\\
University of Amsterdam\\
Amsterdam, The Netherlands\\
\texttt{d.grossi@rug.nl}
}

\maketitle

\begin{abstract}
Power grid operation is becoming more complex due to the increase in generation of renewable energy. The recent series of Learning To Run a Power Network (L2RPN) competitions have encouraged the use of artificial agents to assist human dispatchers in operating power grids. 
However, the combinatorial nature of the action space poses a challenge to both conventional optimizers and learned controllers. Action space factorization, which breaks down decision-making into smaller sub-tasks, is one approach to tackle the curse of dimensionality. In this study, we propose a centrally coordinated multi-agent (CCMA) architecture for action space factorization. In this approach, regional agents propose actions and subsequently a coordinating agent selects the final action. 
We investigate several implementations of the CCMA architecture, and benchmark in different experimental settings against various L2RPN baseline approaches. The CCMA architecture exhibits higher sample efficiency and superior final performance than the baseline approaches. The results suggest high potential of the CCMA approach for further application in higher-dimensional L2RPN as well as real-world power grid settings.
\end{abstract}

\keywords{Multi-Agent Reinforcment Learning (MARL) \and Hierarchical Reinforcement Learning (HRL) \and Consensus-Based Learning \and Power Network Control (PNC)\vspace{5mm}}

\end{@twocolumnfalse}]

\section{Introduction}
The growing demand for electricity, driven by technological advancements and the shift towards electrified industries and transportation, highlights the need for resilient power grids. Managing these grids involves complex sequential decision-making across large state and action spaces, further complicated by aging infrastructure and increased reliance on unpredictable renewable energy sources \cite{mai2018electrification, panciatici2012operating}. 

Traditional grid management methods, such as redispatching and building new infrastructure, are costly and dependent on external factors. However, topological remedial actions, like reconfiguring grid substations, offer a cost-effective alternative that remains underexplored \cite{koglin1982corrective, viebahn2022potential,gridoptions}. 
As power grids expand, conventional computational methods struggle to provide optimal real-time control solutions, leading operators to rely on experience or predefined manuals. The combinatorial complexity of grid configurations necessitates more advanced approaches.

To support research in \emph{power network control} (PNC), the Grid2Op framework was developed \cite{grid2op}, enabling simulation of realistic grid scenarios as a Markov Decision Process (MDP) \cite{sutton2018reinforcement}. 
This framing makes the problem suitable for deep Reinforcement Learning (RL), which has been successful in various domains
\cite{schrittwieser2020mastering, gu2017deep, kendall2019learning, li2019transforming}.

However, deep RL faces scaling issues with large networks due to the combinatorial explosion of state and action spaces. The curse of dimensionality remains a major bottleneck in RL tasks. Specifically, the sample complexity grows exponentially with the dimensionality of the state-action space of the environment, posing challenges for large-scale applications. To address these challenges, Hierarchical Reinforcement Learning (HRL) and Multi-Agent Reinforcement Learning (MARL) offer promising solutions by decomposing complex tasks into simpler sub-tasks and distributing decision-making across multiple agents \cite{dayan1992feudal, littman1994markov}. This idea is captured in the factored MDP framework \cite{osband2014near}, where the original MDP is decomposed into a set of smaller and independently evolving MDPs. In this case, the total sample complexity is determined by the sum of the sizes of the state-action spaces for each individual MDP, rather than their product. As a result, the problem size no longer scales exponentially with the dimension of the problem.

This study integrates the strengths of MARL and HRL to develop a centrally coordinated multi-agent (CCMA) architecture. The design incorporates short-term action suggestions alongside a coordinating agent capable of managing multi-step action sequences. Different algorithms for short-term action suggestions and coordinating agents are compared, ranging from greedy and rule-based strategies to fully RL-based architectures.

The main contributions can be summarized as follows:
\begin{itemize}
    \item We introduce a novel CCMA architecture featuring a coordinator that determines the next action by leveraging regional action suggestions from individual agents alongside global state information. This design enables action space factorization while maintaining centralized coordination. 
    \item We describe several variants of the CCMA architecture, incorporating rule-based, greedy, and RL-based modules. These architectures range from fully deterministic approaches (rule-based and greedy) to fully learned systems (entirely RL-based) and hybrid designs that combine both strategies.
    \item We evaluate the proposed architecture against baseline models \cite{grid2op, l2rpnbaselines}, demonstrating that some of its variants surpass all baselines in both sample efficiency and overall performance, hence effectively factor the MDP. 
    \item Finally, we show that rule-based coordinators are effective in the smallest 5-bus network. However, when tested with larger networks, these architectures no longer match the performance of an RL-based coordinator. 
\end{itemize}

The paper is organized as follows: Section \ref{sec:background} covers the background and related work. Section \ref{sec:con_frame} defines the control framework and Section \ref{sec:methods} describes the power grid environment.
The central Section \ref{sec:power_system_agents} introduces the CCMA architecture.
Finally, Section \ref{sec:results} presents the experiments and results, followed by a discussion and future work in Section \ref{sec:summary}.

\section{Background and Related Work}\label{sec:background}
This section deals with the modeling of the PNC problem (Section \ref{sec:L2RPN}), applications of RL in this domain (Section \ref{sec:RLL2RPN}), and recent HRL and MARL implementations that focus on improving sample efficiency and overcoming the curse of dimensionality (Section \ref{sec:SRL}).

\subsection{Power Network Control}\label{sec:L2RPN}
Secure operation of power networks is required both in normal operating states as well as in contingency states (i.e., after the loss of any single element on the network). That is, the following requirements must be met: (i) In the normal operating state, the power flows on equipment, voltage and frequency are within pre-defined limits in real-time; (ii) In the contingency state the power flows on equipment, voltage and frequency are within pre-defined limits. Loss of elements can be anticipated (scheduled outages of
equipment) or unanticipated (faults for lightning, wind, spontaneous equipment failure). Cascading failures
must be avoided at all times to prevent blackouts (corresponding to the game over state in the RL challenge).

Importantly, each of the control actions performed by power system operators usually not only affects the
current state of the power system but also the future state and availability of future control actions, that
is, short-term actions can have long-term consequences. As a result, the decision problem of power system
operators is typically a sequential decision-making problem in a combinatorial action space in which the current decision can affect all
future decisions. Moreover, due to possible nondeterministic changes of the power system state (e.g., due
to unplanned outages or the intermittent behaviour of renewable energy sources) and different sources of
error (e.g., measurement errors, state estimation errors, flawed judgement) the operators need to handle
uncertainty in their decisions. Finally, operational decisions must often be made quickly, under hard time
constraints \cite{viebahn2022potential}.

\subsection{Deep RL for Power Network Control}\label{sec:RLL2RPN}
Deep RL has demonstrated significant potential in PNC, enabling robust and adaptable behavior over extended time horizons \cite{viebahn2022potential}. This capability surpasses the limitations of expert systems \cite{marot2018expert} and optimization methods, which are constrained by computation time \cite{ruiz2016security,little2021optimal,heidarifar2021optimal}.

In L2RPN competitions, successful solutions typically combine expert rules, RL agents, and brute force simulations for action validation. Notably, RL agents based on Proximal Policy Optimization (PPO) \cite{schulman2017proximal}, with reduced action spaces, have performed well. This approach was used in top-ranking competition entries (e.g. \cite{marot2021learning}, further explored in  studies \cite{lehna2023managing, manczak2023hierarchical, chauhan2023powrl, lehna2024hugo}).

Several commonalities emerge among top-ranking approaches. Most successful methods implement intervention only during hazardous situations \cite{yoon2020winning, lan2020ai, manczak2023hierarchical, chauhan2023powrl, dorfer2022power, lehna2023managing, liu2024progressive}. 
In addition, many such approaches integrate learned modules with simulation-based action evaluation. For instance, this strategy was key to the success of the winning teams in both 2019 \cite{lan2020ai} and 2021 \cite{icapswinner}. 
Notably, the second-place competitor in 2021 employed an advanced expert system, underscoring the value of incorporating domain knowledge from power systems \cite{marot2022learning}. 

\subsection{Curse of Dimensionality}\label{sec:SRL}
The combinatorial complexity of topological actions in PNC poses a challenge for deep RL by hindering complete and consistent convergence and exploration of action and state spaces. HRL simplifies learning by breaking tasks into subtasks, enabling more focused and sample efficient learning per task as well as increased interpretability. MARL distributes decision-making across agents, effectively factorizing the MDP's action space.

\begin{figure*}[t!]
    \centering
    \includegraphics[width=0.9\linewidth]{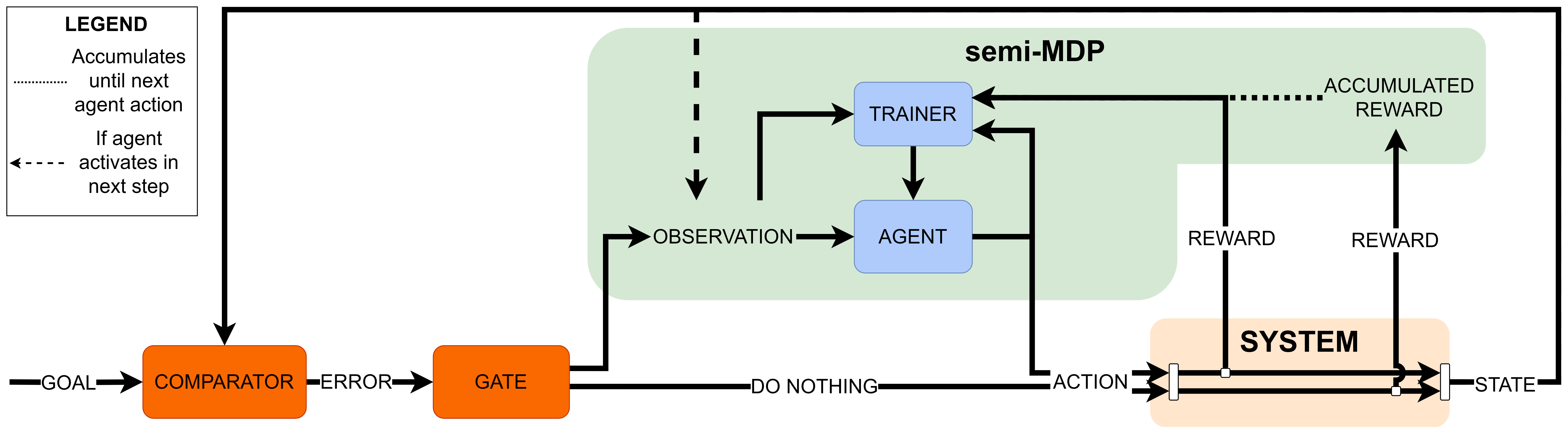}
    \caption{Diagram of the Feedback Control Framework used in this research. The goal to be achieved (and maintained) in the system is an input to the entire framework. The comparator determines the difference between the current state of the system and the desired one. There are then two possible paths for the control flow. If a goal state is reached then no action is performed and the reward is accumulated. If the system is not in the goal state then an agent is called. Periodically this agent can be updated by a trainer. The dashed line represents the only link between the two loops, as the trainer will perceive the accumulated rewards of consecutive time-steps where the {\it do-nothing} was used for control}
    \label{fig:feedback-control}
\end{figure*}

\subsubsection{Hierarchical Reinforcement Learning}
In the L2RPN WCCI 2020 competition, the winning agent for a 36-bus network used a two-tiered framework: a high-level policy proposed a goal topology, while a low-level policy determined the sequence of individual topological actions \cite{yoon2020winning}.

Later, a three-level HRL framework was introduced \cite{manczak2023hierarchical}, where the top level activates the agent in hazardous situations, the intermediate level selects a substation using RL, and the lowest level identifies a configuration. They tested greedy and RL-based approaches for the lowest level, and compared PPO and Soft Actor Critic (SAC) \cite{haarnoja2018soft} for the RL-based modules. They found PPO to have faster convergence, smaller variance, and higher expected rewards.

\subsubsection{Multi-Agent Reinforcement Learning}
The action space in PNC can be factorized into exclusive subsets, one for each substation, enabling a natural MARL framework in a fully cooperative setting where all agents optimize the same collective objective. Following this approach, \citet{van2023multi} extended \citet{manczak2023hierarchical} by introducing multiple agents at the lowest level, with each substation controlled by an agent and a heuristic-based intermediate level. They trained agents using independent PPO (IPPO) and dependent PPO (DPPO). On a 5-bus network, both achieved optimal scores, though single-agent PPO converged faster. However, they anticipate that MARL's advantages will become more pronounced when scaled up to larger networks.

In other domains, \citet{yu2022surprising} showed strong PPO-based multi-agent performance in testbeds, recommending best practices like value normalization, integrating global and local information, 
and limiting sample re-use to avoid training instability from MARL non-stationarity.

\section{Control framework}\label{sec:con_frame}

Figure \ref{fig:feedback-control} shows a schematic of the feedback control (FC) framework employed in this study. 
The \emph{goal} represents a target state that is to be maintained within the controlled \emph{system} (light orange box). The \emph{comparator} (left orange box) computes the \emph{error} between the current and the goal state of the system. The controller produces an \emph{action} to be performed in the system in order to get closer to or maintain the goal state of the system. The system produces a \emph{reward} signal and a new \emph{state} once the action is performed.

As shown in Fig. \ref{fig:feedback-control}, our controller has two modes of operation. After the error is computed, a \emph{gate} (right orange box) determines which mode to use.
If the goal state is achieved ($\text{error}=0$) then the gate selects the lower mode of operation in which a constant action ("do nothing") is performed in the system. If the goal state is not achieved then an \emph{agent} (lower blue box) is used to compute an action. The state of the system (and possibly the error) is used to create the input \emph{observation} for the agent.

The agent can either be rule-based or trained. In case of a trained agent, the agent mode of operation stores past experiences of its interaction with the system for training the agent. The controller can switch between modes of operation multiple times. Switching from the agent to the do-nothing operation may be the result of the agent's action. The system may also remain in the do-nothing mode of operation for multiple timesteps. In this case, the resulting rewards are accumulated. Once the agent mode of operation is used again, the \emph{trainer} (upper blue box) will receive these accumulated rewards.
This allows the agent to receive feedback for the effect that its action has on the gate. Effectively, the agent mode of operation is thus operating as a semi-MDP (sMDP) \cite{SUTTON1999181,Baykal2010}.
The trainer component periodically uses collected experiences to update the parameters of the agent. We note that the trainer does not update the agent every time-stamp. Instead, this happens only periodically.

\section{Power System Environment}\label{sec:methods}
First, in Section \ref{sec:setup}, we describe the power grids used for experimental analysis, which is the \emph{system} in Fig. \ref{fig:feedback-control}. Subsequently, in Section \ref{fcpower}, we describe the corresponding \emph{gate}, \emph{state}, \emph{reward}, and \emph{actions}.

\subsection{Experimental Setup}\label{sec:setup}
We consider two simulated power grids: the \textit{IEEE case 5} and \textit{IEEE case 14} networks. In the Grid2Op simulator \cite{grid2op}, the power grid environments are called \textit{rte\_case5\_example} and \textit{l2rpn\_case14\_sandbox} (see the Grid2Op documentation for visualizations of these power grids). The 5-bus network comprises twenty scenarios, each spanning 2016 timesteps. The 14-bus network contains 1004 scenarios, each unfolding over 8064 timesteps. Each timestep, representing a five-minute interval, incorporates unknown changes in loads and production.

Moreover, unplanned outages can be included. These outages involve temporarily disconnecting a powerline for a specified duration. These unplanned outages are initiated by an opponent, adding stochasticity and further increasing task complexity. This opponent introduces realistic challenges faced by power grid operators, ensuring the system's stability is preserved even during unforeseen contingencies. The opponent is implemented as Grid2Op's \textit{RandomLineOpponent} \cite{randomline}. It is designed to randomly disconnect a line from a predefined set of targets, with a cooldown period implemented to delay consecutive attacks, identical to \citet{manczak2023hierarchical}. After each attack, the disconnected line is automatically restored. Table \ref{tab:opponent_params} outlines the parameters of these attacks. 

\begin{table}[hbt]
    \centering
    \caption{The used parameters for the opponent in Grid2Op, namely the attackable lines ($l_{origin-extremity}$), the attack duration (in timesteps), and frequency (in timesteps).}
    \begin{tabular}{c c c}
    \toprule
     & 5-bus & 14-bus \\
    \midrule
    Attackable lines & $l_{0-2}, l_{0-4}$ & $l_{3-4}, l_{3-6}, l_{3-8}, l_{11-12}$ \\
    Attack duration (ts) & 48 & 48 \\
    Attack cooldown (ts) & 144 & 144 \\
    \bottomrule
    \end{tabular}
    \label{tab:opponent_params}
\end{table}

\subsubsection{Incorporated Domain Knowledge}\label{domainknowledge}
Several domain-inspired features are incorporated in all baselines and architectures to boost performance. First, automatic powerline reconnection is implemented. Powerlines disconnected by overcurrent events face a ten-timestep cooldown before reconnection. Since agents are limited to topological actions, disconnected lines are automatically restored to their previous state one timestep after disconnection. This eliminates the need for the agent to handle reconnections, allowing it to focus exclusively on optimizing grid configurations.

Second, grids tend to be more stable during nighttime. This is due to lower demand (and thus strain) on the grid. This domain knowledge can be exploited to automatically improve the system's performance by
returning the grid back to its original grid state where all powerlines within each substation are interconnected \cite{lehna2023managing}. In this research, this is implemented to happen between 3AM and 6AM. Before this reversion is performed, the agent simulates whether this action would not likely lead to a game-over state, only performing the action if all estimated relative powerline thermal loads remain below 90\% of their maximum capacity.

Lastly, it is worth noting that the grid is typically most stable during nighttime. Normally, agents navigate scenarios ranging from 2016 timesteps (for the 5-bus network) to 8064 timesteps (for the 14-bus), equivalent to seven to 28 days. If an agent fails during training, the scenario terminates. To expose agents to diverse circumstances, scenarios can be subdivided into smaller segments, increasing the likelihood that the agent encounters later timesteps in longer scenarios, potentially speeding up convergence. We therefore split the scenarios into chunks of two days, starting during nighttime. 

\subsection{Power System Feedback Control Elements}\label{fcpower}

This subsection explains how each component of the FC framework presented in Sec. \ref{sec:con_frame} is realized in a power system setting.

\paragraph{Goal}
Obviously, the general goal is to maintain a grid state without any overloaded lines. In this study, we adopt the approach employed in most L2RPN studies. Specifically, the grid state is characterized by the loading of individual power lines, represented as a vector $\rho = (\rho_1, \rho_2, \dots, \rho_n)$ where each element \(\rho_i\) denotes the loading of a specific line in the power grid. The goal state is defined by a threshold \(\tilde{\rho}\), a single scalar value representing the allowable maximum line loading. The goal is to maintain $\max(\rho) < \tilde{\rho}$, where \(\max(\rho)\) refers to the maximum loading among all the lines in the grid. In this study, we set \(\tilde{\rho} = 0.95\).

\paragraph{Gate}
We use a common rule-based gate. If the maximum line loading is lower than the threshold $\tilde{\rho}$ then the do-nothing mode of operation is used. If the maximum line loading surpasses the threshold $\tilde{\rho}$ then the agent mode of operation is used.

\paragraph{State}
The state 
of the system consists of (i) the current grid topology,
(ii) the current load $\rho_{l}$ on each powerline $l$, measured as the fraction of the capacity of the powerline, (iii) the active power flow at each end of each powerline, (iv) the number of overflow timesteps for each powerline, and (v) the current power production and consumption of generators and loads.

In Grid2Op, if a blackout occurs, it results in a game-over state. A blackout happens when the power grid fails to meet the requirements of the generators and loads due to issues like line disconnections, overloading, or other critical failures. Once a blackout is detected, the simulation ends immediately, as the grid's stability can no longer be maintained. 

\paragraph{Reward}

We employ the scaled L2RPN reward \cite{grid2op, manczak2023hierarchical}, which calculates the squared margin of the current flow relative to the current limit for each powerline. The larger the margin, meaning the further the current flow is from the limit, the higher the reward, encouraging states where powerlines are not operating near their maximum capacity.

\paragraph{Actions}

In this study, we introduce a new action space designed to ensure structural N-1 security. Structural N-1 security means that the grid remains operational and avoids an automatic blackout in the event of a single contingency, such as an unplanned powerline outage. We compare the new action space with the default action space which only excludes symmetrical counterparts from all possible substation configurations.

Table \ref{tab:networks} shows the size of each action space.
\begin{table}[tb]
    \centering
    \caption{A comparison between the 5-bus and 14-bus networks. Only topological actions are considered and the do-nothing action is excluded.}
    \begin{tabular}{c c c c c}
    \toprule
     &  \multicolumn{2}{c}{symmetry-filtered} & \multicolumn{2}{c}{N-1 secure} \\
    \midrule
       & \# actions & \# topologies & \# actions & \# topologies \\
    \midrule
    5-bus  & 58 & 31320 & 24 & 364 \\
    14-bus  & 178 & 3.9e+11 & 73 & 3.3e+5 \\
    \bottomrule
    \end{tabular}
    \label{tab:networks}
\end{table}


The size of the symmetry-filtered action space can be computed using two key terms. The first term, \(\alpha(n) = 2^{n-1}\), represents the total number of possible configurations for a substation with two busbars and \(n\) elements connected to it, excluding symmetrical configurations (a factor of \(\frac{1}{2}\)). However, this term does not account for the constraint that no loads or generators (collectively referred to as injections) should be isolated.

The second term, \(\gamma(n') = 2^{n'} - 1\), accounts for all configurations where only injections and no powerlines are connected to a busbar. Here, \(n'\) is the number of injections connected to the substation. The \(2^{n'}\) represents all possible combinations of injections in a configuration where all powerlines are connected to a single busbar, and the \(-1\) excludes the configuration where all injections are also connected to this single busbar.

Using these terms, the size of the symmetry-filtered action space is calculated as:

\begin{equation}\label{eq:alpha}
\tau_{sym} = \alpha(n) - \gamma(n')\ .
\end{equation}

By always maintaining zero or at least two powerlines connected to each busbar, the N-1 secure action space ensures that the failure of a single powerline does not result in an immediate game-over state due to the loss of connection to its injections. 
The size of the structurally N-1 secure action space is:

\begin{equation}\label{eq:TenneTshort}
\tau_{N-1} = \alpha(n) - \gamma(n') - \epsilon(n', n'')\ ,
\end{equation}

\noindent
with:

\begin{equation}\label{eq:epsilon}
\epsilon(n', n'') = 2^{n'}\cdot(n'' - \delta_{n'',2} - \delta_{n'',1})\ ,
\end{equation}

\noindent
where \(n''\) is the number of powerlines, and \(\epsilon(n', n'')\) removes all configurations where only one powerline is connected to a busbar. The Kronecker delta, \(\delta_{n'',k}\), is defined as:

\[
\delta_{n'',k} = 
\begin{cases} 
1 & \text{if } n'' = k, \\
0 & \text{otherwise.}
\end{cases}
\]

\noindent
This means that \(\delta_{n'',2}\) is one when exactly two powerlines are present (\(n'' = 2\)) and zero otherwise, while \(\delta_{n'',1}\) is one when only one powerline exists (\(n'' = 1\)) and zero otherwise. These terms ensure that configurations with invalid powerline setups, such as only one powerline connected or configurations invalid for all possible injection setups, are excluded. In practice, such cases are always kept in the base topology where all elements are connected.

Combined, \(\tau_{N-1}\) excludes all configurations that connect any injection to only a single powerline, ensuring structural N-1 security. This equation is valid as long as at least one powerline exists.

\section{Power system agents}\label{sec:power_system_agents}

In this section, we present the agent configurations used in the second mode of operation of the FC framework (see Sec. \ref{sec:con_frame}, Fig. \ref{fig:feedback-control}). The greedy agent and the single RL agent (Section \ref{sec:greedysingleRL}) approach the sMDP without decomposition.
In contrast, this study introduces a multi-agent architecture that divides the sMDP into a smaller set of independent tasks (see Sec. \ref{sec:architecture}).
The two common baselines serve as a reference to evaluate the impact of decomposing the MDP.

\subsection{Greedy Agent \& Single RL Agent}\label{sec:greedysingleRL}

The greedy-agent baseline is a fully rule-based agent. A greedy agent simulates the effect of every action in the action space for one timestep in the future. Then, the agent selects the action that maximizes a given KPI when simulated. In our case, the KPI is the maximum loading of the grid. See \citet{de2024imitation} for a pseudo-code of the greedy agent.

The single RL agent uses a feed-forward neural network to implement a policy for choosing actions. It also chooses actions from the entire action space. A state-of-the-art RL algorithm is used to train the policy, namely, the Proximal Policy Optimization (PPO) algorithm \cite{schulman2017proximal}.

\subsection{Multi-Agent Architecture}\label{sec:architecture}
\begin{figure}[ht]
    \centering
    \centerline{\includegraphics[width=\linewidth]{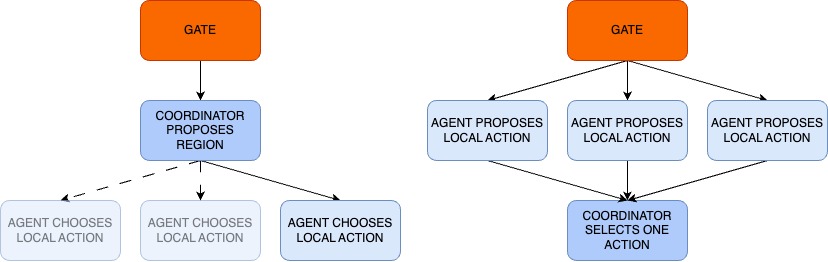}}
    \caption{Left: Hierarchy proposed by \citet{manczak2023hierarchical, van2023multi}. First a region (e.g., a substation) is selected, and then a topological configuration for that region (and that region only) is selected. Right: Hierarchy proposed in this study. For each region a topological reconfiguration is proposed concurrently, and a coordinator then selects the best proposed action (e.g., a substation configuration).}
    \label{fig:difference_hierarchies}
\end{figure}

We divide the overall agent task into two levels of abstraction, following the approaches outlined in \citet{manczak2023hierarchical, van2023multi}. One level consists of deciding the topological reconfiguration of individual regions (e.g. substations). The other level consists of selecting which region should be reconfigured. The agent selecting the region is called the \textit{coordinator}. The reconfiguration of regions may be decided by multiple agents, called \textit{regional agents}, each proposing a topological reconfiguration of its respective region. In our architecture the topological reconfiguration of each individual region is decided first, and the region to be reconfigured is selected second. Figure \ref{fig:difference_hierarchies} illustrates on the left the high-level hierarchy used in \citet{manczak2023hierarchical, van2023multi} where the regional agents are called after the coordinator, and on the right the decision making order proposed in this study. 

Our architecture allows the coordinator to take into account the proposed regional reconfiguration in its decision making process. We believe that this allows for a more clean separation between the two levels of abstraction. In \citet{manczak2023hierarchical, van2023multi}, the coordinator may need to anticipate implicitly what reconfiguration will be proposed for the region it is selecting. This may result in the coordinator having to solve both tasks as a single agent. In other words, we believe that the regional topology selection task needs to be solved (either implicitly or explicitly) before being able to determine which region to select.

\begin{figure*}[tb]
    \centering
    \includegraphics[width=0.9\linewidth]{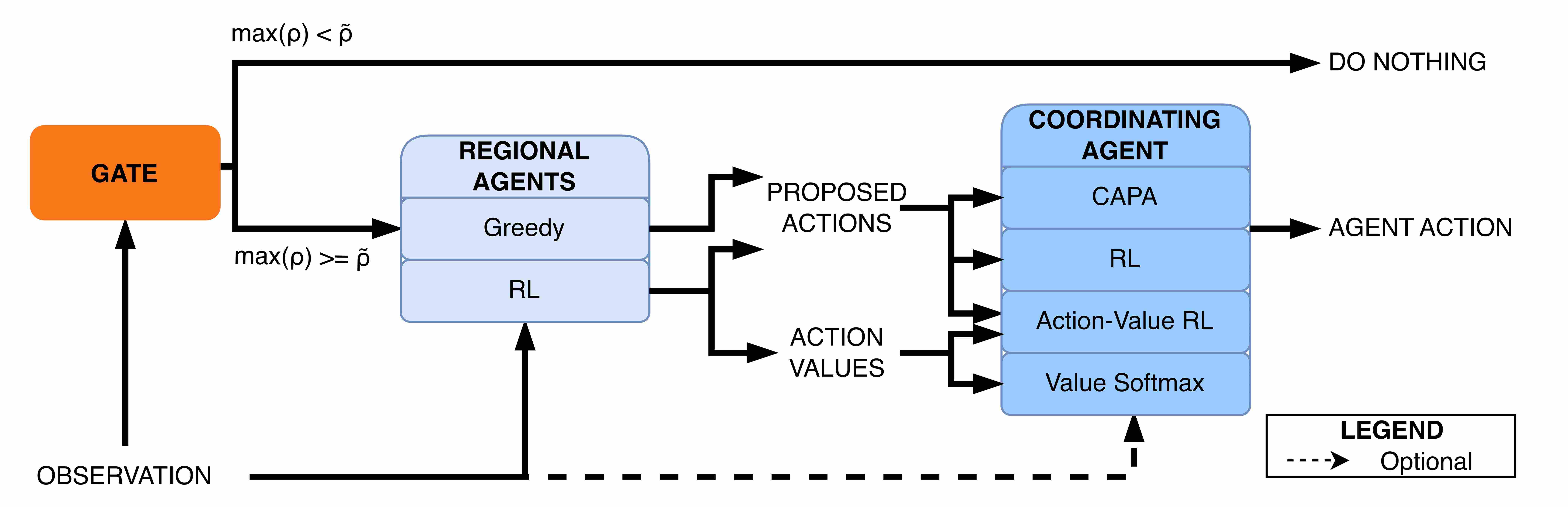}
    \caption{An overview of all possible multi-agent architectures. First the maximum line loading in the current observation \textbf{OBS} is compared to a threshold by the gate to determine whether to not act (upper path) or to try to reconfigure the topology of the grid (lower path). In the latter case, first all the \textbf{REGIONAL AGENTS} propose an action to reconfigure their respective region (or do nothing). This list of action is finally passed to the \textbf{COORDINATING AGENT} which select one of the regions (i.e. one of the actions proposed by any of the regional agents). The final result is always an action in the topological reconfiguration space.}
    \label{fig:agent-architecture}
\end{figure*}

Figure \ref{fig:agent-architecture} provides a more detailed description of our architecture. At the top level, there is always the FC framework's gate.
When switching to the agent mode of operation, the regional agents are activated. Each of them receives the current observation and subsequently proposes a topological reconfiguration for its region\footnote{A regional agents may also still propose a \textit{do-nothing} action for its respective region. }. All of the proposed regional reconfigurations are passed forward to the coordinator. The latter may take into account all of the proposed actions and the current observation. If available, the coordinator also receives the action-value estimates, which represent the network's internal state and quantify the predicted utility or confidence in the effectiveness of each action.

Each regional agent only experiences rewards (and thus transitions) if it is selected by the coordinator. When the gate selects consecutive do-nothing actions, the accumulated reward is assigned to the last regional agent that was selected by the coordinator.

Figure \ref{fig:agent-architecture} also shows the different implementations of the regional agents and the coordinator considered in this study. In this paper, each regional agent is either a RL agent trained using the PPO algorithm \cite{schulman2017proximal}, or a greedy rule-based method (see Sec. \ref{sec:greedysingleRL}). 
For the RL agent, the action-value of the proposed action (as evaluated by the critic) may also be passed forward. 

The coordinator also has multiple implementations.
When the coordinator is trained via RL and the action-value of the regional agent is included then the coordinator is referred to as \textit{Action-Value RL} agent. If the action-value is not passed forward, the coordinator is simply referred to as \textit{RL} agent.
We also experiment with two rule-based implementations of the coordinator. The \textit{CAPA heuristics} coordinator uses the \textit{CAPA} algorithm \cite{yoon2020winning}. Algorithm \ref{alg:modified_capa} shows our version of the CAPA coordinator, which looks at the actions proposed by the regional agents only to exclude those who proposed do-nothing actions. Moreover, this agent does have memory for the purpose of exhausting a priority list between regions before generating a new one. 
The other rule-based coordinator selects a regional agent based solely on the action-value of their actions\footnote{Note that these action-values are proposed by independent critics, one for each regional agent.}.
More precisely, the \textit{Value Softmax} coordinator samples a regional agent's action with a probability proportional to their action-value in order to balance exploitation of the action-values once the critics are sufficiently trained. This also allows for exploration thus avoiding biasing training data only towards specific regional agents.

\begin{algorithm}[h]
\caption{Ordered CAPA Algorithm}
\label{alg:modified_capa}
\begin{algorithmic}[1]
\STATE Given observation $obs$, activation threshold $max\_rho$ and an initial priority list of actions $P$ that acts as the state of the algorithm
\STATE $current\_rho \gets$ $max(rho)$ in $obs$
\IF{$P = None$ or $P$ is empty}
    \STATE Generate priority list $P$ using CAPA policy
\ENDIF
\FOR{each action $A$}
    \IF{$A \neq do\_nothing$}
        \STATE Remove $A$ from $P$
        \STATE \textbf{return} $A$ and $P$
    \ENDIF
\ENDFOR
\STATE $P \gets None$
\STATE \textbf{return} $do\_nothing$ and $P$
\end{algorithmic}
\end{algorithm}

We note that if the \textit{coordinating agent} uses a rule-based method that takes into account only the current observation to select a region (to then simply return whatever action that region's agent has proposed) then we effectively also obtain the architecture shown in Fig. \ref{fig:difference_hierarchies} \textbf{(left)}. This is because, in such a case, it is irrelevant whether or not the regional agents propose the actions before or after the coordinating agent.

Table \ref{tab:comparison-complexity} in Appendix \ref{ap:comparelit} highlights the key differences between the proposed CCMA architecture and other state-of-the-art architectures in power grid topology control.

\subsection{Overview of power system agents}\label{sec:agents_overview}

We summarize all considered approaches using the following naming convention: The full multi-agent system is indicated by [name regional agent]-[name coordinating agent], with the names given as in Table  \ref{tab:input_output_architectures} and Fig. \ref{fig:agent-architecture}.

\begin{table}[h]
\centering
\caption{The input for each module in the agent architecture.}
\begin{tabular}{ll}
\toprule
Agent Name & Input \\ 
\midrule
\multicolumn{2}{c}{Regional Agents} \\ 
\midrule
Greedy & Observation \\
RL & Observation \\
\midrule
\multicolumn{2}{c}{Coordinating Agents} \\ 
\midrule
CAPA & Regional Actions \& Observation \\
RL & Regional Actions \& Observation \\
Action-Value RL & Regional Action-Value Pairs \& Observation \\
Value Softmax & Regional Action-Value Pairs \\ 
\bottomrule
\end{tabular}
\label{tab:input_output_architectures}
\end{table}

\subsubsection{Baselines.}
To evaluate the proposed approach, results are compared with several baselines that should at least be met in terms of performance:
\begin{itemize}
    \item \textbf{Do-Nothing Agent}: This baseline shows the number of timesteps survived when no action is taken at any timestep.
    \item \textbf{Single RL Agent}: This baseline uses a single PPO agent responsible for configuring the entire network. The agent receives network observations as input and outputs a single action. 
    \item \textbf{RL-CAPA}: This architecture combines a rule-based CAPA coordinator with RL-based regional agents. The CAPA policy prioritizes substations at risk of immediate overloading, using the highest relative overload value for selection. This approach ensures that regional agents are trained in scenarios where their actions are likely to have a significant impact. This adapted CAPA policy based on \citet{van2023multi} also considers past actions to avoid experience imbalance and repeated selection of the same substation. If a proposed action does not alter the grid topology, the next highest-priority substation proposes an action, as detailed in Alg. \ref{alg:modified_capa}.
\end{itemize}

\subsubsection{Benchmarks.}

Two benchmark agents are designed using action simulation and greedy strategies. Both architectures are computationally intensive, as they simulate all possible actions and select the optimal one in a greedy manner. However, they serve to illustrate the complexity within networks and scenarios, demonstrating the potential scope of learning.

\begin{itemize}
    \item \textbf{Greedy Agent}: This baseline involves a greedy agent that makes decisions based on minimizing the maximum relative line loading of the network. It operates without hierarchical coordination and provides a benchmark of what a simple architecture with enough computational power can achieve.
    \item \textbf{Greedy-CAPA}: To test the rule-based potential of a hierarchical design, greedy regional agents are paired with the CAPA coordinator.
\end{itemize}

\subsubsection{Multi-agent RL architectures.}
For the proposed CCMA architecture several implementations are explored:

\begin{itemize}
    \item \textbf{Greedy-RL}: This setup features a hierarchical structure with an RL-based coordinator that receives proposals from regional agents to address congestion. Unlike a greedy agent, the RL coordinator benefits from implicit planning, similar to the \textit{PPO Substation} implementation by \citet{manczak2023hierarchical}.
    The key distinction is that \citet{manczak2023hierarchical}'s implementation does not involve multiple regional agents or consider the greedy agent’s planned actions.
    \item \textbf{RL-RL}: This architecture features both a learned coordinator and learned regional agents. The coordinator utilizes both global state observations and regional agent information, inspired by the approach suggested by \citet{yu2022surprising}. This setup allows the system to learn multi-step actions but may introduce instability due to the interactions among multiple learned levels and agents.
    \item \textbf{RL-Action Value RL}: This approach extends the previous setup by incorporating the action-value function outputs from regional agents as additional input for the learned coordinator. This allows for a more refined decision-making process by integrating the value estimates of different actions, potentially enhancing the coordination and overall performance of the system.
    \item \textbf{RL-Value Softmax}: This architecture uses the action-value function outputs from regional agents. A coordinator employing 
    \textit{softmax} (see Equation \ref{eq:softmax} below) assigns probabilities to actions during training, promoting exploration and fair experience distribution. During evaluation, \textit{argmax} is used to choose the optimal action based on the trained agents.
\end{itemize}


\subsection{Implementation details}
To improve learning, avoid overfitting, and ensure a fair comparison between architectures, several measures are implemented for each experiment, including train-validation-test splitting, scenario splitting, and identical hyperparameter tuning.

\subsubsection{Trainer}
For all experiments, PPO \cite{schulman2017proximal} is used for learning a policy. The policy, denoted by $\pi$, is parameterized using a neural network with learnable parameters $\theta$. The action-value function, represented as \(Q(s_t, a_t)\), and the state-value function \(V(s_t)\), are represented by a neural network with parameters $\phi$.

For both training and evaluation, PPO is employed as a stochastic policy, meaning actions are sampled from a probability distribution rather than being chosen deterministically. Common practices such as including Generalized Advantage Estimation (GAE) are adopted to reduce variance and improve training stability \cite{yu2022surprising}. The PPO implementation by \citet{liang2017ray} is utilized in this research.

For the multi-agent implementation, it is common that one agent acquires significantly more experience than other agents. To counteract this, we adapted PPO to postpone training until all agents have reached the minimum batch size or until any agent has reached double the minimum batch size.

\subsubsection{Machine Learning Practices}
Hyperparameter tuning is critical in RL research and applications because hyperparameters greatly affect the performance, stability, and efficiency of RL algorithms. For each architecture and environment, two grid searches are conducted to tune the parameters using Ray Tune \cite{liaw2018tune}. The first is a coarse grid search to identify suitable learning rates, batch sizes, minibatch sizes, and the number of iterations. The second grid search is narrower, focusing on smaller variations in these parameters, as well as additional parameters such as the clipping range, value function coefficient range, and GAE lambda range. Appendix \ref{ap:grid_search} details the full grid search settings for all experiments.

It should be noted that to validate the stability and reliability of the proposed method, all architectures are trained independently per network using ten different random seeds. The stability of the approach is illustrated explicitly in Figure \ref{fig:training_curves}, where the variability across multiple seeds is represented by the standard deviation, indicating the robustness and convergence behavior under diverse initialization conditions.

Additionally, a priority function is integrated into the selection of the next training scenario to accelerate learning \cite{yoon2020winning, van2023multi}. This function assigns a logit to each scenario based on its characteristics, as defined by Equation \ref{eq:prio}:

\begin{equation}
\label{eq:prio}
w_{prio} = 1 - \Big(2 \cdot \sqrt{\frac{t_{survived}}{t_{max}}}\Big) \,
\end{equation}

\noindent
where \(w_{prio}\) represents the logits, \(t_{survived}\) is the number of timesteps the agent survived in the last iteration, and \(t_{max}\) is the maximum number of timesteps for that scenario. The logits are scaled between -1 and 1: fully completed scenarios are assigned a value of -1, scenarios that fail at the start are assigned a value of 1, and unseen scenarios are initialized with a logit of 2 to encourage exploring new scenarios first.
These logits are then used in a softmax distribution (Eq. \ref{eq:softmax}) to sample the next scenario:

\begin{equation}\label{eq:softmax}
\sigma(x_{i}) = \frac{e^{x_i}}{\sum_j e^{x_j}}\ ,
\end{equation}

\noindent
where \(\sigma(x_{i})\) represents the probability of selecting scenario $i$, and the denominator ensures the probabilities are normalized by summing the exponential values of all scenario logits.

To avoid overfitting and ensure a robust model, all scenarios are divided into a 70\%/10\%/20\% train/validation/test split.

\section{Results}\label{sec:results}
\begin{figure*}[t]
     \centering
     \includegraphics[width=0.78\linewidth]{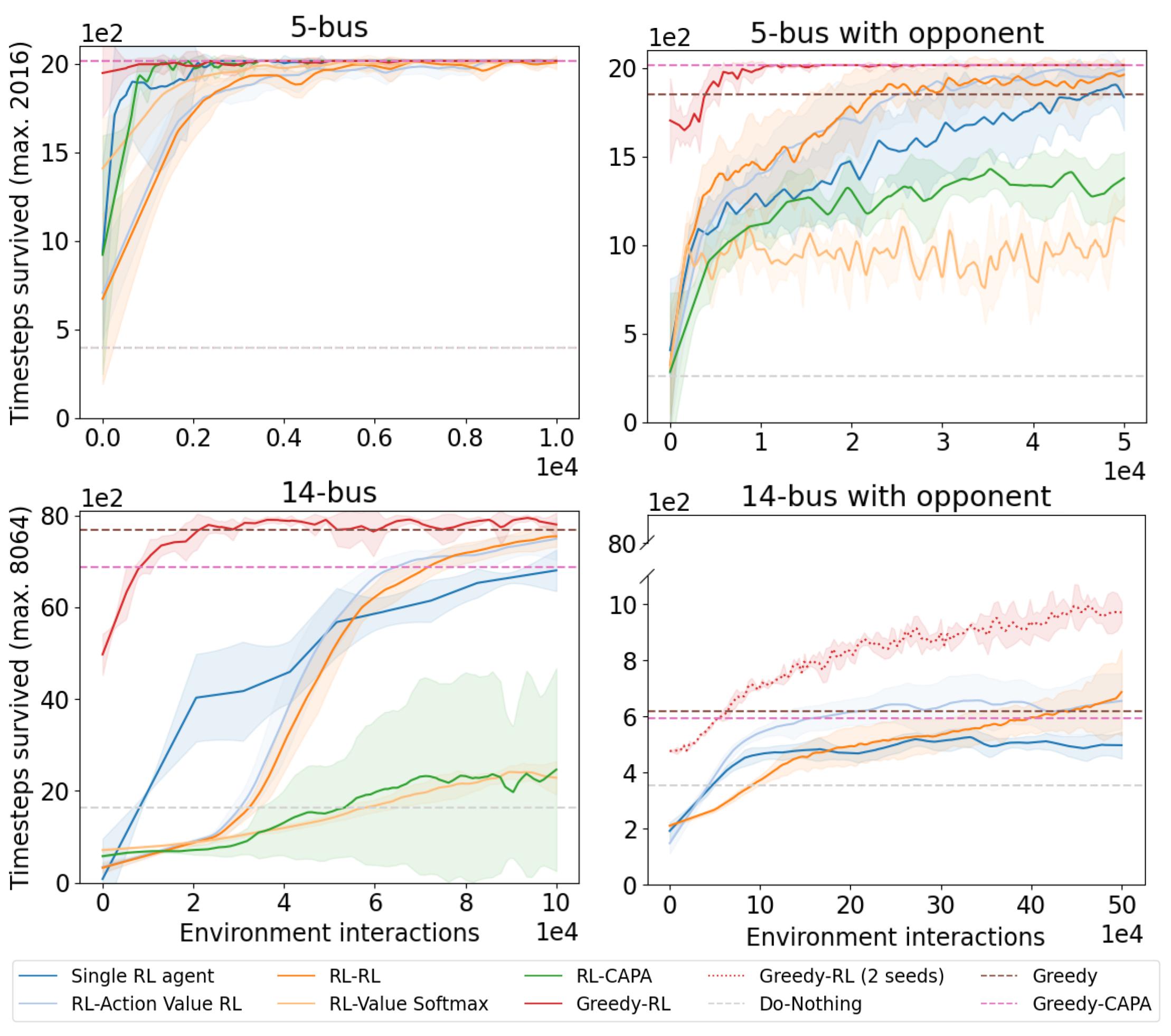}
     \caption{The mean number of timesteps survived on all validation scenarios throughout training for the 5-bus network without opponent (top left), the 5-bus network with opponent (top right), the 14-bus network without opponent (bottom left) and the 14-bus network with opponent (bottom right). For the mean and standard deviation, seeds are interpolated between 0 and \texttt{max\_timesteps} to align them on a common time axis.}
     \label{fig:training_curves}
\end{figure*}

Figure \ref{fig:training_curves} shows the learning curves of the different agent architectures for the four different experimental setups, namely, the agent performances during training on all validation scenarios, averaged over all ten model seeds\footnote{It should be noted that the training curves for the 14-bus network without an opponent have not yet fully converged. Due to the computational complexity involved in grid searching and generating all models across multiple seeds, training was halted once the curves had leveled off at 100,000 timesteps.}.

First, we note that all agent architectures considered in this study, except the do-nothing agent, achieve optimal performance in the 5-bus environment without opponent (upper-left panel in Fig. \ref{fig:training_curves}, first column in Table \ref{tab:combined_performance}). Consequently, the 5-bus environment without opponent can be considered as a baseline (or code testing) environment which is used to demonstrate that 
the minimally expected learning behavior is exhibited. Hence, in the subsequent sections the results related to the 5-bus environment without opponent are not discussed in detail. 

\subsubsection*{Rule-based agents}
In this study, three entirely rule-based agents are considered (see Sec. \ref{sec:agents_overview}).
The do-nothing agent (dashed-grey in Fig. \ref{fig:training_curves}) represents a baseline agent which is completely inflexible.
The greedy agent (dashed-purple in Fig. \ref{fig:training_curves}) and the Greedy-CAPA agent (dashed-brown in Fig. \ref{fig:training_curves}) represent benchmark approaches which include a high amount of simulation.
Correspondingly, the do-nothing agent shows the worst performance of all considered approaches for all environments. For the 5-bus environment without opponent (with opponent) it survives about $10\%$ ($8.5\%$ on average) of the possible timesteps, whereas for the 14-bus network it survives $18\%$ ($4.3\%$ on average).
In contrast, the two rule-based benchmark approaches show high performances for both the 5-bus environment with opponent (upper-right panel in Fig. \ref{fig:training_curves}) and the 14-bus environment without opponent (lower-left panel in Fig. \ref{fig:training_curves}). More precisely, in the 5-bus environment with opponent the Greedy-CAPA agent achieves $100\%$ of the possible timesteps, whereas the greedy agent achieves $83\%$ on average.
In the 14-bus environment without opponent the order is reversed with the greedy agent achieving $96\%$ of the possible timesteps, whereas the Greedy-CAPA agent achieves $85\%$. The 14-bus environment with opponent (lower-right panel in Fig. \ref{fig:training_curves}) is significantly more challenging than the previous environments. Consequently, the achieved timesteps of the greedy agent and the Greedy-CAPA agent are significantly smaller with only $7.7\%$ and $7.4\%$ on average, respectively.

\subsubsection*{Single RL baseline agent}\label{sec:performance_PPOsingle}
The single RL agent (blue in Fig. \ref{fig:training_curves}) represents the most common RL-based baseline approach for L2RPN \cite{marot2021learning}. For the 5-bus environment with opponent (upper-right panel in Fig. \ref{fig:training_curves}) and the 14-bus environment without opponent (lower-left panel in Fig. \ref{fig:training_curves}), it exhibits performances similar to the rule-based benchmark approaches (i.e. the greedy and Greedy-CAPA agents).
However, in the most challenging environment (14-bus with opponent, lower-right panel in Fig. \ref{fig:training_curves}) its performance drastically decreases 
and remains clearly below the already low performances of the two rule-based benchmark approaches.

\subsubsection*{Low-performing trained multi-agent systems}\label{sec:performance_low}
RL-CAPA (green in Fig. \ref{fig:training_curves}) and RL-Value Softmax (light-orange in Fig. \ref{fig:training_curves}) represent the two multi-agent architectures in which the regional agents are trained via RL, whereas the coordinating agent is rule-based (see Sec. \ref{sec:agents_overview}).
Both approaches show significantly worse performances than all other investigated agent architectures (except the do-nothing agent) for all environments (except the baseline 5-bus environment without opponent). In particular, the single RL baseline agent is outperforming RL-CAPA and RL-Value Softmax in both the 5-bus environment with opponent (upper-right panel in Fig. \ref{fig:training_curves}) 
and the 14-bus environment without opponent (lower-left panel in Fig. \ref{fig:training_curves}). 
Due to these very low performances we decided to not train these approaches in the significantly more challenging and computationally expensive 14-bus environment with opponent.

\subsubsection*{High-performing trained multi-agent systems}\label{sec:performance_high}
Greedy-RL (red in Fig. \ref{fig:training_curves}),
RL-RL (orange in Fig. \ref{fig:training_curves}), and RL-Action Value RL (light-blue in Fig. \ref{fig:training_curves}) represent the multi-agent architectures in which the coordinating agent is trained via RL (see Sec. \ref{sec:architecture}). The RL-RL multi-agent system and the RL-Action Value RL multi-agent system show very similar results such that we focus on RL-RL in the remainder of this section. All three approaches outperform the single RL baseline in the challenging 14-bus environments (lower panels in Fig. \ref{fig:training_curves}). 
The learning curves of RL-RL are initially below the learning curves of the single RL baseline but start to excel after a moderate amount of training. 
Importantly, RL-RL eventually also outperforms the rule-based benchmark approaches (i.e., the greedy agent and Greedy-CAPA)
in the most challenging 14-bus environment with opponent (lower-right panel in Fig. \ref{fig:training_curves}).

The most impressive sample efficiency and performance is exhibited by Greedy-RL. In each of the four panels in Fig. \ref{fig:training_curves}, the training curve of Greedy-RL starts at a significantly higher level than those of the other trained agents. This is related to the employment of greedy agents as the regional agents such that the coordinating agent can only choose from valuable actions. Subsequently, Greedy-RL continues to exhibit the highest performance level during the entire training period in each environment.
For the 5-bus environment with opponent (upper-right panel in Fig. \ref{fig:training_curves}) and the 14-bus environment without opponent (lower-left panel in Fig. \ref{fig:training_curves}), Greedy-RL converges quickly, effectively solving the environment while the training curves of the other trained agents only start to exhibit learning progress.

Also in the most challenging 14-bus environment with opponent (lower-right panel in Fig. \ref{fig:training_curves}), Greedy-RL exhibits extraordinary sample efficiency and performance. The learning curves of Greedy-RL start at the performance level of the converged single RL baseline models. We note that for Greedy-RL the learning curves related to different seeds, shown in Figure \ref{fig:compare_single} in Appendix \ref{individual_seeds}, exhibit bimodal behavior (in contrast to all other architectures).
That is, one group of seeds remains at a lower performance level (but still higher than the level of the converged single RL baseline) whereas a second group of seeds exhibits steep learning curves outperforming both the single RL baseline  and the rule-based benchmark approaches (i.e., the greedy agent and Greedy-CAPA) by a large margin. Hence, a mean curve (averaged across all seeds) is not a good representation of the model behavior such that in the lower-right panel of Fig. \ref{fig:training_curves} we only show an average across high-performing seeds.
More details on the training curves related to individual seeds can be found in App. \ref{individual_seeds}.

For completeness, Table \ref{tab:combined_performance} in Appendix \ref{ap:test_results} shows model evaluation on the test scenarios, which exhibits results consistent with the evaluation on the validation scenarios during training. However, the focus of this study is on the \emph{sample efficiency} of agent architectures, and, hence, in this study the model evaluation during training (Fig. \ref{fig:training_curves}) is more relevant than final test results (Tab. \ref{tab:combined_performance} in App. \ref{ap:test_results}).

\section{Discussion }\label{sec:summary}
From the experiments it is evident that there can be benefits to factoring the action space. The Greedy-RL, RL-RL, and RL-Action Value RL agents consistently outperform the single RL agent, both asymptotically and in sample efficiency. With the larger grid, and especially with contingencies, this pattern becomes more evident.

There seems to be no significant difference in performance between the RL-RL and RL-Action Value RL agents. This suggests that the value of the regional agents' critics does not hold useful information for solving the coordination task.

On the other end, the Greedy-RL agent significantly outperforms RL-RL agent, both asymptotically and in sample efficiency. The improved sample efficiency could be due to the warm-start that the Greedy-RL agent can benefit from, as regional agents are already fixed. In contrast, agents like RL-RL may suffer from non-stationarity since both levels need to perform reasonably well before they can learn appropriately. Concretely, it can lead to poor performance when a regional agent selects a good action but a coordinator does not select the appropriate agent, or vice versa.

On the other hand, the Greedy-RL agent exhibits bi-modal behavior in the stochastic environment (Fig. \ref{fig:compare_single} in App. \ref{individual_seeds}). One possible explanation is that not all contingency cases have equal value. It could be that only a smaller subset of the contingency cases pushes the agent to pick safer actions. This may have happened only on certain seeds. One possible explanation as to why we observe this only with Greedy-RL is its higher sample-efficiency. This would make the agent faster at picking up such patterns when (and if) they appear.

These results suggest that it is crucial to have long term reasoning in the coordination task. With an RL coordinator the Greedy-RL agent manages to outperform the Greedy baseline, especially in the challenging stochastic setting, despite the regional Greedy agents optimizing topological configurations for a single timestamp. The importance of long term reasoning in coordination is further supported by the low asymptotic performance of RL-CAPA and RL-Value Softmax. For the simplest setting, a 5-bus network without opponent, the coordinators such as CAPA-heuristic, random substation selection, Value Softmax still perform reasonably well. However, when used with added complexity due to an adversary or a larger network, these architectures no longer match the performance of a learned coordinator.

Moreover, when the network grows in size, the number of possible substations to select also grows. This makes the coordinator's task increasingly complex, calling the need for a more sophisticated coordinator.

Finally, because the Greedy-RL is among the best performing agents, it suggests that the factored action spaces can be tackled independently.
In practice, the Greedy-RL agent is trained as a single RL agent within its coordination task. 
Thus, the regional agents can be trained independently of the coordinator. The coordinator can also be trained individually, but only after the regional agents.
Most likely, this results in the significantly higher sample-efficiency of the Greedy-RL agent.

As a final note, Fig. \ref{fig:training_curves} illustrates the stability of the proposed method by illustrating the standard deviation over ten different random seeds and multiple validation scenarios used during training. It stands out that some architectures exhibit considerably lower variability across validation scenarios and seeds, as reflected by smaller standard deviations, indicating more stable and robust convergence behavior. In contrast, other architectures show higher variability, suggesting that their performance is more sensitive to initialization.

\section{Conclusion and Future Work}\label{sec:summary2}
This study introduces a new approach that combines HRL with MARL to tackle the growing complexity of PNC in increasingly unpredictable environments. Our three-level architecture effectively breaks down the decision-making process, making control strategies more modular. One notable aspect is the introduction of a coordinating agent that considers both global state information and action proposals from regional agents.

\subsection{Future Work}
Future research could build on this study by exploring several avenues for extending and refining the proposed methods. Testing the scalability and robustness of the approach in larger networks, such as the 36-bus \cite{marot2022learning} and 118-bus networks \cite{serre2022reinforcement}, could provide insights into its effectiveness in more complex environments. To assess scalability, we conducted a preliminary evaluation on a 36‑bus network (66,810 possible topological actions) and find promising outcomes. 
The results are shown in Appendix \ref{ap:36}. Moreover, future studies could test the impact of different strategies for decomposing the actions space \cite{polimifactorization}.

Another promising direction is to investigate cooperative mechanisms among agents. While this study used a fully independent setup with PPO, sharing parameters or using a shared critic could potentially stabilize learning and enhance efficiency. Testing the impact of partial observability on regional agents could also provide more realistic insights into local versus global information. 




The results also suggest more concrete next steps for making the CCMA more scalable. The Greedy-RL architecture shows the most promise in sample efficiency. However, it is also the most computationally expensive agent as detailed in Appendix \ref{ap:compute_time}. To mediate this, one could use imitation learning to train regional agents. These agents would try to imitate the greedy agent directly just much faster.


Furthermore, the regional agents can be trained independently of the coordinator using RL, similar to a single-agent RL setting where the action space is confined to a single regional agent. This approach allows regional agents to incorporate some level of longer-term planning, albeit with a shorter time horizon compared to the coordinator that could be limited to a few timesteps. The aforementioned imitation learning can be used as warm start in this setting.


However, it should be noted that our approach does not factorize the observation space. With the current observation space, when training the regional agents separately, the data should include a lot of varied configurations of the observed section of the grid. This applies also when using only imitation learning. Each regional agent observes the entire grid, and thus all other regions. Therefore the amount of data needed to train an individual agents scales exponentially with the number of regions. 

One could try to give partial observations to the regional agents. Each agent only taking as observation a neighborhood of its region (included). Thus, the data to train a regional agents should only include combinations within its neighborhood. This effectively creates a cap on the amount of data needed to train an agent, with respect to the number of regions. In such a setting, the amount of data scales with respect to the size of the neighborhood and not the total number of regions. Expanding the grid outside of a regional agent's neighborhood will not affect the amount of data needed to train it. The connections going out of the neighboring regions can still be kept and simulated. The data can include various level of loading on these connections. Effectively the part outside of the neighborhood can be seen as grid injections. It should be noted that these observation spaces would not be completely independent as there would be overlap between them.

\section*{Acknowledgments}
We want to thank Erica van der Sar and Alessandro Zocca from the Vrije Universiteit Amsterdam for their valuable insights while conducting the research as well as feedback in refining this work.

AI4REALNET has received funding from European Union’s Horizon Europe Research and Innovation programme under the Grant Agreement No 101119527. Views and opinions expressed are however those of the authors only and do not necessarily reflect those of the European Union. Neither the European Union nor the granting authority can be held responsible for them.

Davide Grossi acknowledges support by the \href{https://hybrid-intelligence-centre.nl}{Hybrid Intelligence Center}, a 10-year program funded by the Dutch Ministry of Education, Culture and Science through the Netherlands Organisation for Scientific Research (NWO).

\bibliographystyle{ACM-Reference-Format}

\bibliography{biblio}

\appendix
\newpage
\onecolumn

\section{Comparative Analysis of Power Grid Control Architectures}\label{ap:comparelit}

Table \ref{tab:comparison-complexity} compares the proposed CCMA architecture and other state-of-the-art architectures, focusing on agent architecture, action space representation and action space reduction.

\begin{table*}[ht]
\caption{Comparison of RL-based Architectures for Power Grid Topology Control.}
\centering
\footnotesize
\begin{tabular}{|p{2cm}|p{5.5cm}|p{4.5cm}|p{4.5cm}|}
\hline
\textbf{Paper} & \textbf{Agent Architecture} & \textbf{Type of Action Space for RL Agent} & \textbf{Action Space Reduction} \\
\hline

\textbf{CCMA, proposed in this study} & 
Hierarchical multi-agent architecture that separates decision layers between, first, regional agents and, second, a coordinator.
Distributed proposals over multiple regional agents yield a diverse range of candidate actions. Directly addresses the combinatorial challenge by decomposing the action space, so the problem scales with the sum of sub-problem sizes rather than their product.
&
RL coordinator learns on higher-level action space consisting of the set of controllable substations. This action space is naturally much smaller than the primitive action space.
& 
No arbitrary action space reduction is necessary. Only power system principles are used to reduce the action space, without relying on heuristics or manual tuning. This also avoids introducing bias through greedy action selection.
\\
\hline

\textbf{\citet{liu2024progressive}} & 
Hierarchical agent architecture similar to \citet{manczak2023hierarchical}, as discussed in Sec. \ref{sec:architecture} in this study. That is, first a coordinator specifies a substation, and subsequently a single regional agent specifies the substation configuration.
&
As in CCMA, the coordinator learns on higher-level action space consisting of the set of controllable substations. However, the coordinator does not receive the output of the regional agent as input, as discussed in Sec. \ref{sec:architecture} in this study.
& 
Similar to CCMA, that is, no action space reduction but action space decomposition. However, no action space decomposition on the subsation level.
\\
\hline

\textbf{\citet{yoon2020winning}} & 
Hierarchical approach consisting of, first, a RL agent choosing a goal topology, and, second, a rule-based agent to specify the corresponding sequence of primitive substation changes. Moreover, an afterstate representation is used.
&
RL agent learns on entire topology action space which, hence, needs to be drastically reduced.
&
The action space consists of goal topologies, which suffers even more from combinatorial explosion than substation reconfiguration actions. How the action space is reduced is not discussed in the paper, but the code includes an arbitrary cut-off range of maximum possible actions per substation.
\\
\hline

\textbf{\citet{lehna2023managing}} & 
Employs a teacher-tutor-junior-senior framework. Essentially, this is a single RL approach with a warm-start based on greedy expert data.
& 
RL agent needs to learn on the primitive action space which, hence, needs to be drastically reduced.
& 
Uses brute-force greedy search to create a frequency distribution of the entire action space. Subsequently, the action space is then drastically pruned based on frequency distribution using an arbitrary cut-point.
\\
\hline

\textbf{\citet{lehna2024hugo}} & 
Extending \citet{lehna2023managing} by adding a rule-based application of so-called target topologies in a low loading regime.
& 
Same as \citet{lehna2023managing}, that is, RL agent needs to learn on the primitive action space which, hence, needs to be drastically reduced.
& 
Similar to \citet{lehna2023managing}. Reduced set of target topologies is based greedy agent statistics. The reduced set of primitive actions is simply taken from another agent by \citet{dorfer2022power}.
\\
\hline

\textbf{\citet{chauhan2023powrl}} & 
Essentially a single RL approach supported by a number of domain specific heuristics similar to \citet{lehna2023managing}, see also Sec. \ref{domainknowledge} in this study.
&
Same as \citet{lehna2023managing}, that is, RL agent needs to learn on the primitive action space which, hence, needs to be drastically reduced.
& 
Similar to \citet{lehna2023managing}. Through extensive brute-force simulation, the action space is drastically reduced; the actual action space reduction procedure is not explained.
\\
\hline

\textbf{\citet{dorfer2022power}} &
Essentially a single RL approach using Monte Carlo Tree Search.
&
Same as \citet{lehna2023managing}, that is, RL agent needs to learn on the primitive action space which, hence, needs to be drastically reduced.
&
Similar to \citet{lehna2023managing}. Through extensive brute-force simulation, the action space is drastically reduced by using the top most frequent actions.
\\
\hline

\end{tabular}
\label{tab:comparison-complexity}
\end{table*}

\section{Hyperparameters}\label{ap:grid_search}
A grid search was performed for all architectures in all different set-ups, of which the explored values can be found here. Unless otherwise mentioned, the default parameters of Ray's PPO were used \cite{liang2017ray}.

\subsection{5-bus network without opponent}
Grid search was performed over the following parameters:
\begin{itemize}
    \item Learning rate (lr): 0.01, 0.001, 0.0001, 0.00001
    \item SGD minibatch size (mbs): 32, 64
    \item Train batch size (bs): 64, 128
\end{itemize}

Depending on the result of the grid search, the learning rate parameter was further explored to include half and five times its best setting. For example, if 0.001 was the best learning rate, the next grid search also explored 0.005 and 0.0005. Additionally, the number of SGD iterations (it) was fixed at 15. Table \ref{tab:param5reg} illustrates all chosen parameters.

\begin{table}[H]
    \centering
    \caption{Chosen parameters for all learning architectures on the 5-bus network without opponent.}
    \begin{tabular}{rllll}
        \toprule
        Agent & lr & mbs & bs & it\\
        \midrule
        Single RL & 0.01 & 32 & 64 & 15 \\
        RL-RL & 0.0005 & 64 & 128 & 15 \\
        RL-Action Value RL & 0.0005 & 64 & 128 & 15 \\
        \midrule
        RL-Action Softmax & 0.0005 & 64 & 128 &  15 \\
        RL-CAPA & 0.005 & 32 & 64 & 15 \\
        RL-Random & 0.01 & 64 & 64 & 15 \\
        Greedy-RL & 0.0005 & 32 & 128 & 15 \\
        \bottomrule
    \end{tabular}
    \label{tab:param5reg}
\end{table}

\subsection{5-bus network with opponent}
Grid search was performed over the following parameters:
\begin{itemize}
    \item Learning rate (lr): 0.001, 0.0001, 0.00001, 0.000001
    \item SGD minibatch size (mbs): 32, 64, 128, 256
    \item Train batch size (bs): 128, 256, 512, 1024
    \item Number of SGD iterations (it): 5, 10
\end{itemize}
Depending on the result of the grid search, the learning rate parameter was further explored to include nearby values. For example, if 0.001 was the best learning rate, the next grid search also explored 0.0005, 0.0007, 0.003, 0.005. Table \ref{tab:param5opp} illustrates all chosen parameters.

\begin{table}[H]
    \centering
    \caption{Chosen parameters for all learning architectures on the 5-bus network with opponent.}
    \begin{tabular}{rllll}
        \toprule
        Agent & lr & mbs & bs & it \\
        \midrule
        Single RL & 0.0007 & 128 & 512 & 10 \\
        RL-RL & 0.0007 & 64 & 128 & 5 \\
        RL-Action Value RL & 0.0003 & 128 & 256 & 5 \\
        \midrule
        RL-Value Softmax & 0.0005 & 64 & 128 & 10 \\
        RL-CAPA & 0.0005 & 128 & 256 & 5 \\
        RL-Random & 0.0001 & 64 & 128 & 5 \\
        Greedy-RL & 0.00005 & 64 & 256 & 10 \\
        \bottomrule
    \end{tabular}
    \label{tab:param5opp}
\end{table}

\subsection{14-bus network without opponent}
Grid search for the 14-bus networks was performed in two batches. The first batch was a coarse grid search over the following settings:
\begin{itemize}
    \item Learning rate (lr): 0.0001, 0.00001, 0.000001
    \item SGD minibatch size (mbs): 64, 256, 1024
    \item Train batch size (bs): 128, 512, 2048
    \item Number of SGD iterations (it): 5, 10, 15
\end{itemize}

The second batch further investigated the learning rate, SGD minibatch size and train batch size parameters. For example, if 0.001 was the best learning rate, the next grid search also explored 0.0005, 0.0007, 0.003, 0.005. For the batch sizes, the surrounding integers $x$ in $2^{x}$ were explored. Additionally, new parameters were explored:
\begin{itemize}
    \item Value function clip parameter (vfc): 10, 20, 50
    \item Clip parameter (cp): 0.15, 0.2, 0.3
    \item Value function loss coefficient (vfl): 0.9, 0.95, 1
    \item Lambda ($\lambda$): 0.9, 0.92, 0.95, 0.98, 1
\end{itemize}

Table \ref{tab:param14reg} illustrates all chosen parameters.

\begin{table*}[h!]
    \centering
    \caption{Chosen parameters for all learning architectures on the 14-bus network without opponent.}
    \begin{tabular}{rllllllll}
        \toprule
        Agent & lr & mbs & bs & it & cp & vfc & vfl & $\lambda$ \\
        \midrule
        Single RL & 0.00005 & 256 & 1024 & 15 & 0.3 & 10 & 1 & 0.95 \\
        RL-RL & 0.00007 & 512 & 256 & 10 & 0.15 & 10 & 1 & 0.95 \\
        RL-Action Value RL & 0.0001 & 256 & 512 & 10 & 0.15 & 20 & 0.95 & 0.95 \\
        \midrule
        RL-Value Softmax & 0.00005 & 64 & 256 & 10 & 0.15 & 10 & 1 & 0.92 \\
        RL-Capa & 0.00005 & 128 & 256 & 10 & 0.3 & 10 & 1 & 1 \\
        RL-Random & 0.00001 & 256 & 256 & 10 & 0.3 & 10 & 1 & 0.92 \\
        Greedy-RL & 0.0005 & 128 & 512 & 15 & 0.3 & 10 & 0.95 & 0.95 \\
        \bottomrule
    \end{tabular}
    \label{tab:param14reg}
\end{table*}

\subsection{14-bus network with opponent}
Grid search for the 14-bus networks was performed in two batches. The first batch was a coarse grid search over the following settings:
\begin{itemize}
    \item Learning rate (lr): 0.0001, 0.00001, 0.000001
    \item SGD minibatch size (mbs): 64, 256, 1024
    \item Train batch size (bs): 128, 512, 2048
    \item Number of SGD iterations (it): 5, 10, 15
\end{itemize}

The second batch further investigated the learning rate, SGD minibatch size and train batch size parameters. For example, if 0.001 was the best learning rate, the next grid search also explored 0.0005, 0.0007, 0.003, 0.005. For the batch sizes, the surrounding integers $x$ in $2^{x}$ were explored. Additionally, new parameters were explored:
\begin{itemize}
    \item Value function clip parameter (vfc): 10, 20, 50
    \item Clip parameter (cp): 0.15, 0.2, 0.3
    \item Value function loss coefficient (vfl): 0.9, 0.95, 1
    \item Lambda ($\lambda$): 0.9, 0.92, 0.95, 0.98, 1
\end{itemize}

Table \ref{tab:param14opp} illustrates all chosen parameters.

\begin{table*}[h!]
    \centering
    \caption{Chosen parameters for all learning architectures on the 14-bus network with opponent.}
    \begin{tabular}{rllllllll}
        \toprule
        Agent & lr & mbs & bs & it & cp & vfc & vfl & $\lambda$ \\
        \midrule
        Single RL & 0.0005 & 512 & 2048 & 5 & 0.15 & 10 & 0.95 & 0.95 \\
        RL-RL & 0.0001 & 1024 & 1024 & 10 & 0.15 & 10 & 0.95 & 0.95 \\
        RL-Action Value RL & 0.0001 & 1024 & 2048 & 10 & 0.15 & 10 & 0.95 & 0.95 \\
        \midrule
        Greedy-RL & 0.0001 & 1024 & 1024 & 10 & 0.2 & 10 & 1 & 0.95 \\
        \bottomrule
    \end{tabular}
    \label{tab:param14opp}
\end{table*}

\section{Individual Seeds for 14-Bus Network With Opponent}\label{individual_seeds}

\begin{figure*}[h]
    \centering
    \begin{subfigure}[b]{0.45\textwidth}
        \includegraphics[width=\textwidth]{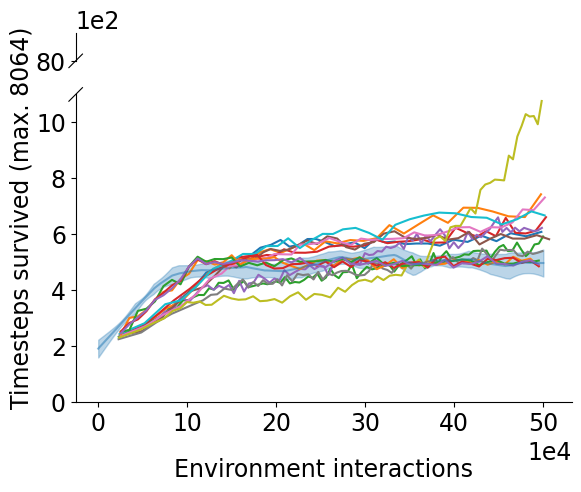} 
        \caption{The individual seeds for the RL-RL Agent compared to the mean and standard deviation of the Single RL Agent (blue).}
        \label{fig:subfig1}
    \end{subfigure}
    \hfill
    \begin{subfigure}[b]{0.45\textwidth}
        \includegraphics[width=\textwidth]{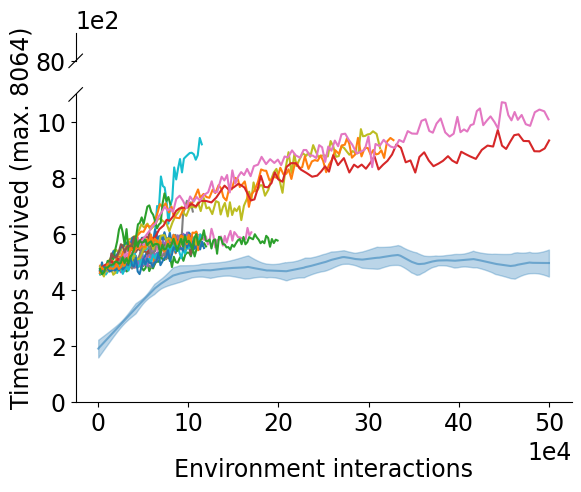} 
        \caption{The individual seeds for the Greedy-RL Agent compared to the mean and standard deviation of the Single RL Agent (blue).}
        \label{fig:subfig2}
    \end{subfigure}

    \caption{The mean performance of the Single RL Agent compared to the individual seeds of the Greedy-RL and RL-RL agent on the 14-bus environment with opponent. For the mean and standard deviation, seeds are interpolated between 0 and \texttt{max\_timesteps} to align them on a common time axis.}
    \label{fig:compare_single}
\end{figure*}

Figure \ref{fig:compare_single}
shows the mean performance of the Single RL agent and compares it to the individual seeds of the RL-RL agent (left) and the Greedy-RL agent (right) on the 14-bus environment with opponent. 
The training of the Greedy-RL agent is very computationally expensive due to the simulation of the entire action space and the stochastic nature of the environment (Sec. \ref{fcpower}). We were able to complete training for the 14-bus environment with opponent only on two of the ten seeds. However, we still observe that it consistently outperforms the average performance of the single RL agent, both asymptotically and in sample efficiency.

\section{Results of test scenarios}\label{ap:test_results}
Table \ref{tab:combined_performance} shows the number of time steps survived in the test scenarios. For each architecture, the results of best-performing model seed in the validation scenarios is shown.

\begin{table*}[h!]
    \centering
    \caption{Performance of different agents (best performing model seed on validation scenarios) in the test scenarios of the 5-bus and 14-bus networks, with and without opponent, in terms number of timesteps survived. Shown are the mean and standard deviation (sd) across different scenarios (and different environment seeds in case of opponent).}
    \begin{tabular}{rllll}
        \toprule
        Architecture
        & \multicolumn{2}{c}{\# ts survived (sd) in 5-bus network} & \multicolumn{2}{c}{\# ts survived (sd) in 14-bus network} \\
        & Regular & Opponent & Regular & Opponent \\
        \midrule
        Do-Nothing & 212.8 (208.5) & 172.0 (187.6) & 1527.9 (1343.3) & 349.5 (326.7) \\
        Single RL & 2016.0 (0.0) & 2016.0 (0.0) & 6744.4 (2289.7) & 516.8 (468.4) \\
        RL-CAPA & 2016.0 (0.0) & 1514.3 (747.1) & 1533.6 (1353.8) & - \\
        \midrule
        Greedy & 2016.0 (0.0) & 1662.3 (664.8) & 7716.8 (1305.6) & 622.4 (500.0) \\
        Greedy-CAPA & 2016.0 (0.0) & 2016.0 (0.0) & 6841.8 (2109.3) & 593.6 (434.5) \\
        \midrule
        Greedy-RL & 2016.0 (0.0) & 2016.0 (0.0) & 7853.8 (952.0) & 923.7 (893.5) \\
        RL-Value Softmax & 2016.0 (0.0) & 1270.2 (799.5) & 2316.0 (1987.9) & - \\
        RL-RL & 2016.0 (0.0) & 1977.6 (242.5) & 7510.7 (1600.5) & 1122.4 (1033.6) \\
        RL-Action Value RL & 2016.0 (0.0) & 1976.7 (248.7) & 7410.0 (1531.0) & 835.6 (764.7) \\
        \bottomrule
    \end{tabular}
    \label{tab:combined_performance}
\end{table*}

\section{Preliminary 36-bus Experiment}\label{ap:36}

To explore scalability on the 36‑bus network (\textit{l2rpn\_icaps\_2021\_large}), we experimented with the Greedy, Single RL and Greedy-RL agents.

First, we deployed a more efficient greedy agent that evaluates all non‑hub substations (fewer than 1,000 actions) and executes the first move whose simulated peak line load remains below the safety threshold or decreases by at least 5\%, thus avoiding full combinatorial search over hub actions. Only if no non‑hub action qualifies, the agent continuously samples from the 65,121 hub actions until a good action is found. If still none meet the criteria, it selects the action minimizing the maximum line loading. 

Next, we built a reduced action space by collecting every hub action chosen more than once by this greedy agent (while retaining all non‑hub actions). We used this action space to train four and two sets of hyperparameters of both the Greedy‑RL and Single‑RL agents, respectively. Based on App. \ref{ap:grid_search}, we evaluated the parameter setting specified in Table \ref{tab:hyperparams_agents_36bus}.

\begin{table*}[h]
  \centering
  \caption{Hyperparameter configurations for Greedy‑RL and Single‑RL on the 36‑bus network without opponent.}
  \label{tab:hyperparams_agents_36bus}
  \begin{tabular}{rllllllll}
    \toprule
    Agent       & lr               & mbs   & bs     & it   & cp   & vfc & vfl  & $\lambda$ \\
    \midrule
    Greedy‑RL   & 0.0001 & 512   & 3072   & 10   & 0.1  & 20   & 0.95 & 0.95 \\
    Greedy‑RL   & 0.00001 & 512   & 3072   & 10   & 0.1  & 20   & 0.95 & 0.95\\
    Greedy‑RL   & 0.00005 & 1024  & 3072   & 5    & 0.2  & 20   & 0.95  & 0.95\\
    Greedy‑RL   & 0.00005 & 1024  & 3072   & 5    & 0.2  & 20   & 0.95  & 0.95\\
    \midrule
    Single‑RL   & 0.00005 & 512   & 3072   & 10   & 0.1  & 20   & 0.95  & 0.95\\
    Single‑RL   & 0.00005 & 512   & 3072   & 10   & 0.1  & 20   & 0.95  & 0.95\\
    \bottomrule
  \end{tabular}
\end{table*}

\begin{figure*}[h] 
  \centering
  \begin{subfigure}[t]{0.48\linewidth}
    \includegraphics[width=\linewidth]{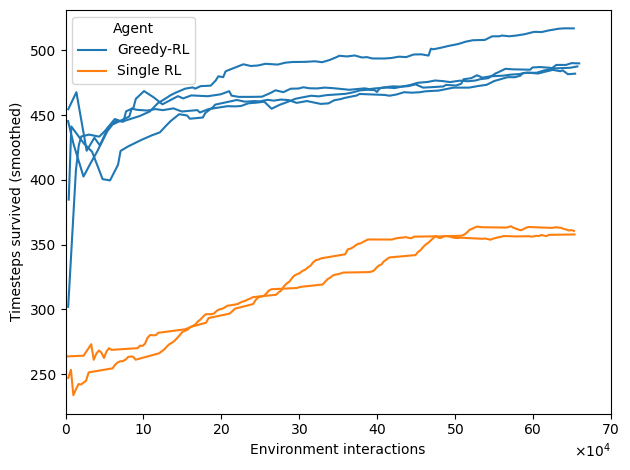}
    \caption{The rolling mean (window = 50) number of timesteps survived on training scenarios for the 36-bus network without opponent. As described in Sec. \ref{domainknowledge}, the agent is, whenever feasible, trained on two‑day segments (576 timesteps) extracted from the original training scenarios. When nighttime instabilities prevent a full two‑day run, these segments may occasionally be one day shorter or longer. Across all training scenarios, the average segment length is 588.1 timesteps.}
    \label{fig:36train}
  \end{subfigure}
    \hfill
  \begin{subfigure}[t]{0.48\linewidth}
    \includegraphics[width=\linewidth]{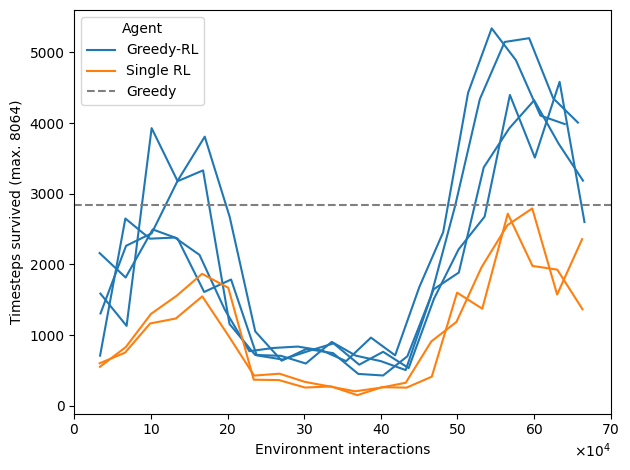}
    \caption{The mean number of timesteps survived on all 296 validation scenarios throughout training for the 36-bus network without opponent.}
    \label{fig:36initial}
  \end{subfigure}
  \caption{Performance of the Greedy, Greedy-RL and Single RL agents on the 36-bus network without opponent over time: (a) training, (b) validation.}
  \label{fig:36_combined}
\end{figure*}

Figure \ref{fig:36_combined} shows the resulting training (a) and validation (b) curves. The results show that both the Single RL and Greedy-RL suffer from overfitting, suggesting that further hyperparameter tuning is needed. While the training curves consistently increase (see Fig. \ref{fig:36train}), the validation curves show large variations (see Fig. \ref{fig:36initial}). It should be noted that when an agent is updated, its buffer will contain experiences from only a fraction of the training scenarios. One likely explanation for this overfitting (and the subsequent recovery) is that out of distribution training scenarios are picked for training (at around 200k environmental interactions in Fig. \ref{fig:36initial}). This most likely results in the agent trying to perform well in these peculiar scenarios, effectively ``forgetting" some of the behaviors useful in the validation set.

Regardless, the plots still highlight the value of Greedy-RL in terms of performance. The best validation performance achieved by Greedy-RL far surpasses that of both the Single-RL and the rule-based Greedy agent. These findings confirm our smaller‑grid conclusions while underscoring the need for a full hyperparameter grid search. Future work will also explore splitting the hub agent into multiple regional agents, allowing the full range of actions to be used again.

\section{Computational Time}\label{ap:compute_time}

Table \ref{tab:compute_times} presents the average computational time per environment interaction during training for various agents on the 14-bus network with an opponent, using a single CPU per worker. The Single RL agent demonstrates the lowest computational time, followed closely by the RL-RL and RL-Action Value RL agents. In contrast, the Greedy-RL agent exhibits significantly higher computational time, approximately an order of magnitude greater than the others.

\begin{table*}[h]
    \centering
    \caption{Compute Time Per Timestep While Training 14-Bus Network With Opponent}
    \label{tab:compute_times}
    \begin{tabular}{c|c}
    Agent & Time per Environment Interaction (s) \\
    \hline
        Single RL & 0.307 \\
        RL-RL & 0.492 \\
        RL-Action Value RL & 0.468 \\
        Greedy-RL & 5.089\\
    \end{tabular}
\end{table*}

\end{document}